\documentclass[aps,prl,reprint,superscriptaddress,floatfix]{revtex4-2}
\usepackage{hyperref}

\usepackage[T1]{fontenc}
\usepackage[utf8]{inputenc}
\usepackage{lmodern}      
\usepackage{amsmath,amssymb,amsthm}
\usepackage{bm}          
\usepackage{physics}      
\usepackage{graphicx}    
\usepackage{microtype}    
\usepackage{xcolor}
\graphicspath{{figs/}}

\begin{document}

\title{Magnon-Mediated Superconductivity in the Infinite-$U$ Triangular Lattice}

\author{Hantian Zhu}
\affiliation{Department of Physics and Astronomy, University of Tennessee, Knoxville, Tennessee 37996, USA}

\author{Yixin Zhang}
\affiliation{Department of Physics and Astronomy, University of Tennessee, Knoxville, Tennessee 37996, USA}

\author{Shang-Shun Zhang}
\affiliation{Department of Physics and Astronomy, University of Tennessee, Knoxville, Tennessee 37996, USA}

\author{Yang Zhang}
\email{yangzhang@utk.edu}
\affiliation{Department of Physics and Astronomy, University of Tennessee, Knoxville, Tennessee 37996, USA}
\affiliation{Min H. Kao Department of Electrical Engineering and Computer Science, University of Tennessee, Knoxville, Tennessee 37996, USA}

\author{Cristian D. Batista}
\email{cbatist2@utk.edu}
\affiliation{Department of Physics and Astronomy, University of Tennessee, Knoxville, Tennessee 37996, USA}
\affiliation{Neutron Scattering Division, Oak Ridge National Laboratory, Oak Ridge, TN, USA}

\begin{abstract}
We demonstrate that the infinite-$U$ triangular-lattice Hubbard model supports a superconducting state built from tightly bound Cooper pairs composed of two holes and one magnon ($2h1m$). Building on the seminal prediction of repulsively bound $2h1m$ states, we show that next-nearest-neighbor hopping $t_{2}$ coherently mixes symmetry-related configurations, stabilizing an $s$-wave bound state with substantial binding energy and a light effective mass. Large-scale DMRG calculations at finite doping identify a magnetization plateau corresponding to a gas of such bound states and quasi--long--range superconducting order with power-law $2h1m$ pair correlations. Our results establish a magnon-mediated superconducting mechanism driven by kinetic frustration, with immediate detectable signatures for moiré Hubbard materials and ultracold-atom simulators.
\end{abstract}

\maketitle
A central challenge in condensed matter physics is to understand how unconventional superconductivity (SC) can emerge from purely short-range \emph{repulsive} interactions, as exemplified by the Hubbard model~\cite{Hubbard1963,Kanamori1963,Gutzwiller1963,Arovas2022HubbardReview,LeeNagaosaWen2006}. While conventional approaches invoke the exchange of virtual spin fluctuations to mediate pairing between fermions, here we explore a more radical mechanism: the formation of composite bosons in which magnetic excitations are physically bound to charge carriers. In this scenario, mobile carriers strongly reorganize the surrounding magnetic background, and the resulting many-body dressing can generate emergent effective attractions, even in the absence of any bare attractive interaction.

Recent advances in ultracold-atom quantum simulators and transition-metal dichalcogenide (TMD) moir\'e heterostructures have turned the Hubbard model into a highly tunable experimental platform~\cite{Wu_PRL_2018,Zhang_PRB_2020,Regan_Nature_2020,Tang_Nature_2020,Li_Nature_2021,Xu_NatNano_2022,Jordens_Nature_2008,Mazurenko_Nature_2017,Greif_Science_2013}. These systems have enabled direct observation of long-standing theoretical predictions, including Nagaoka-type kinetic ferromagnetism and antiferromagnetic AFM spin-polaron formation~\cite{Nagaoka1966,HaerterShastryPRL2005,SposettiPRL2014,PhysRevB.107.224420,ZhangZhuBatista2018,doi:10.1126/sciadv.adp5681}. In square-lattice realizations, a single doped carrier produces a pronounced distortion of AFM correlations, forming a magnetic ``cloud''~\cite{Koepsell_Nature_2019}. Upon doping a magnetic insulator, the magnon excitations themselves are likewise strongly renormalized by itinerant carriers, giving rise to magnetic polarons~\cite{Prichard_NatPhys_2025}.

Frustration further enriches this physics. In the triangular-lattice Hubbard model at large $U/t$, hole doping enhances antiferromagnetic correlations~\cite{Haerter2005}, whereas electron doping produces extended ferromagnetic ``bubbles''~\cite{Prichard_Nature_2024,Xu2023FrustrationDoping,Lebrat2024NagaokaPolaron}. This pronounced particle--hole asymmetry originates from the fact that single-hole motion is \emph{kinetically frustrated} on a ferromagnetic background, while single-electron motion is not. The frustration can be locally relieved when the two spins forming a triangle with the hole bind into a singlet state. This simple mechanism not only accounts for kinetically induced antiferromagnetism on frustrated lattices, but also underlies a growing class of unconventional states in doped Mott insulators, including spin polarons~\cite{ZhangZhuBatista2018,ZhangFu2023}, resonating-valence-bond states on corner-sharing lattices~\cite{Glittum2025}, and frustration-driven pairing phenomena~\cite{ZhangZhuBatista2018,doi:10.1126/sciadv.adp5681,zhang2025antiferromagnetismtightlyboundcooper}.

In triangular moir\'e heterostructures such as MoTe$_2$/WSe$_2$, recent optical studies~\cite{Tao2024SpinPolaron,Ciorciaro_Nature_2023} have observed the predicted bound state of a hole and a flipped spin (relative to full spin polarization) with total spin $S=3/2$~\cite{ZhangZhuBatista2018}. In this spin-polaron state, the magnon localizes on a triangle shared with the hole and forms a singlet with the neighboring spins, effectively reversing the sign of the local hopping amplitude (equivalent to a $\pi$ flux) and thereby relieving kinetic frustration. A particularly clear experimental signature of a Fermi gas of such $S=3/2$ polarons is the emergence of a magnetization plateau at $M=(1-3\delta)/2$, corresponding to one spin flip per doped hole, as observed in moir\'e bilayers~\cite{Tao2024SpinPolaron} and in quantitative agreement with theoretical predictions~\cite{ZhangFu2023}.

As shown in Ref.~\cite{ZhangZhuBatista2018}, this mechanism can be extended further: a single magnon can bind two holes into a composite Cooper pair when carrier motion is sufficiently frustrated, for instance by including a third-neighbor hopping $t_3$. Subsequent theoretical work demonstrated that a second-neighbor hopping $t_2$ provides an even more effective route to stabilizing the two-hole--one-magnon ($2h1m$) bound state in the triangular-lattice Hubbard model~\cite{doi:10.1126/sciadv.adp5681}. This three-body composite can be viewed as a spin polaron that captures an additional hole, providing a concrete example of how the coupling between charge motion and local magnetism can generate effective attraction from purely repulsive interactions. These results naturally raise the question: \emph{Can moir\'e or ultracold-atom realizations of the Hubbard model be tuned into a regime where such $2h1m$ pairs proliferate and condense, giving rise to a magnon-mediated superconducting phase?}

Using density matrix renormalization group (DMRG) simulations at finite hole doping $\delta$~\cite{White_PRL_1992_DMRG}, we demonstrate the stability of a dilute gas of $2h1m$ quasiparticles. This regime is signaled by the emergence of a magnetization plateau at $M_p=(1-2\delta)/2$, corresponding to one spin flip per pair of holes, and is characterized by quasi--long--range superconducting order of the composite Cooper pairs. While $2h1m$ pair--pair correlations decay as a power law, correlations of bare two-hole pairs remain short-ranged, unambiguously establishing that the magnon is not merely a mediator of attraction but an essential constituent of the Cooper pair itself.

We consider the infinite-$U$ Hubbard model on a triangular lattice with nearest-neighbor (NN) hopping $t_{1}$ and next-nearest-neighbor (NNN) hopping $t_{2}$ in the presence of an external magnetic field $\bm{H} = H \hat{\bm z}$:
\begin{equation}
\mathcal{H} = -t_{1} \!\!\sum_{\langle ij\rangle \sigma} \!\! c^{\dagger}_{i\sigma} c_{j\sigma}
-t_{2}\!\!\!\sum_{\langle \langle ij \rangle \rangle \sigma} \!\!\! c^{\dagger}_{i\sigma} c_{j\sigma}+ U \!\sum_{i} n_{i\uparrow} n_{i\downarrow} - h \sum_{i} S^{z}_{i},
\end{equation}
where $h = g \mu_B H$, $\mu_B$ is the Bohr magneton, $g$ is the gyromagnetic factor and $\langle ij \rangle$ ($\langle \langle  ij \rangle \rangle$) denotes NN (NNN) bonds $ij$.

\begin{figure}[t] %
  \centering
  \includegraphics[width=1\columnwidth]{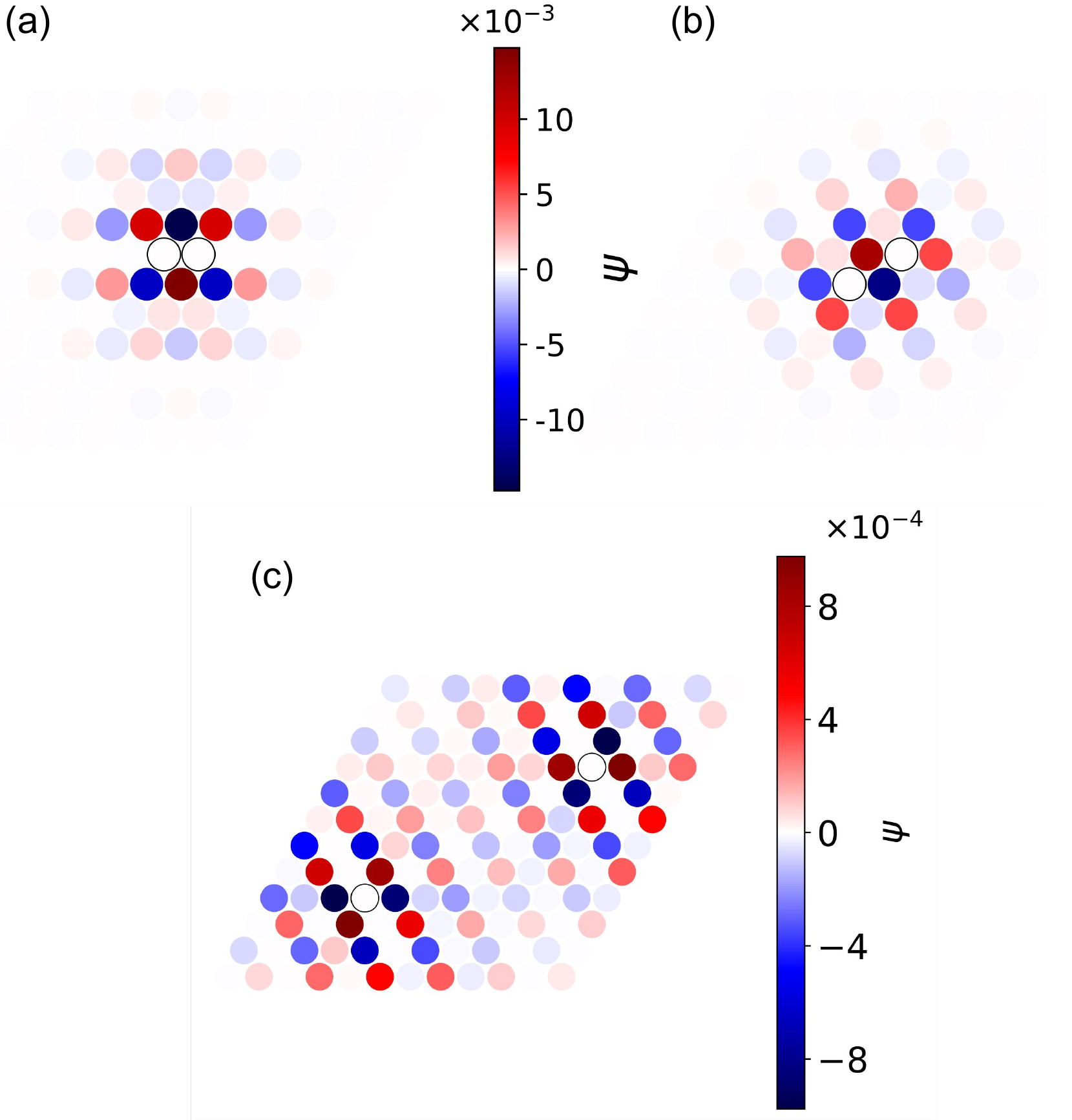} %
  \caption{Ground-state wave function from $12\times12$ ED for $t_1=1,t_2=0.3$ with the two hole positions
fixed at: (a) a NN bond; (b) a NNN
bond; (c) the maximal separation allowed by periodic boundary conditions.}
  \label{fig:1}
\end{figure}

We begin by considering the two-hole problem in the magnetic-field window where the ground state has 
$S^{z}\equiv \sum_i S^z_i = N/2 - 2$, corresponding to a fully polarized background with a single spin flip. 
Building on the analysis of Nazaryan and Fu~\cite{doi:10.1126/sciadv.adp5681}, we revisit the two-hole--one-magnon ($2h1m$) bound state, which appears over a finite field range for $t_{2}/t_{1} \gtrsim 0.09$. 
For $t_{2}/t_{1} \gtrsim 0.15$, two such quasiparticles repel each other, indicating the stability of a dilute gas of $2h1m$ composites. 
Motivated by these results, we set $t_{1}=1$ and $t_{2}=0.3$ and perform exact diagonalization on a $12\times12$ lattice with periodic boundary conditions in both directions. 
The ground-state wave function in the $2h1m$ sector is obtained using the \texttt{QuSpin} package~\cite{Weinberg2017}. 
We then analyze the spatial structure of the bound state by fixing the positions of the two holes and evaluating the wave function as the magnon is moved across the lattice:
\begin{equation}
\psi(k\,|\,i,j) =
    \langle \Psi_G |\, c_{j\uparrow} c_{i\uparrow} S_k^- \,| \mathrm{FM}\!\uparrow \rangle\\
\quad S_k^- = c_{k\downarrow}^\dagger c_{k\uparrow}
\label{eq:phi_r_lj}
\end{equation}
Here $k$ denotes the position of the mobile magnon, and $i,j$ are the fixed positions of the two holes. The ground-state wave functions for three representative hole configurations are shown in Fig.~\ref{fig:1}. As the hole--hole separation increases from nearest neighbor to next-nearest neighbor  and then to the farthest distance on the lattice, the magnon probability distribution evolves from strongly localized to progressively more delocalized, while the overall magnitude $\lvert\psi(k; i,j)\rvert$ decreases. This trend indicates the formation of a tightly bound three-body pair in which the two holes and the magnon cluster closely together.

To characterize the three-body bound state more explicitly, we introduce a real-space pair-creation operator that generates a $2h1m$ configuration on the fully polarized reference state $\ket{\mathrm{FM}\!\uparrow}$:
\begin{equation}
\hat{\Delta}_{\hat{\mathbf{r}}_1,\hat{\mathbf{r}}_2,(x,y)}
\equiv
c_{(x,y)+\hat{\mathbf{r}}_1\,\uparrow}\,
c_{(x,y)+\hat{\mathbf{r}}_2\,\uparrow}\,
S^{-}_{(x,y)},
\label{eq:Delta_ijk_def}
\end{equation}
where $(x,y)$ denotes the magnon position and $\hat{\mathbf{r}}_{1,2}$ are the displacement vectors from the magnon to the two holes. By rotating this operator on the triangular lattice and tracking the phase acquired under $C_{6}$ transformations, we find that the bound state transforms trivially and therefore possesses $s$-wave symmetry.

\begin{figure}[t] %
  \centering
  \includegraphics[width=1\columnwidth]{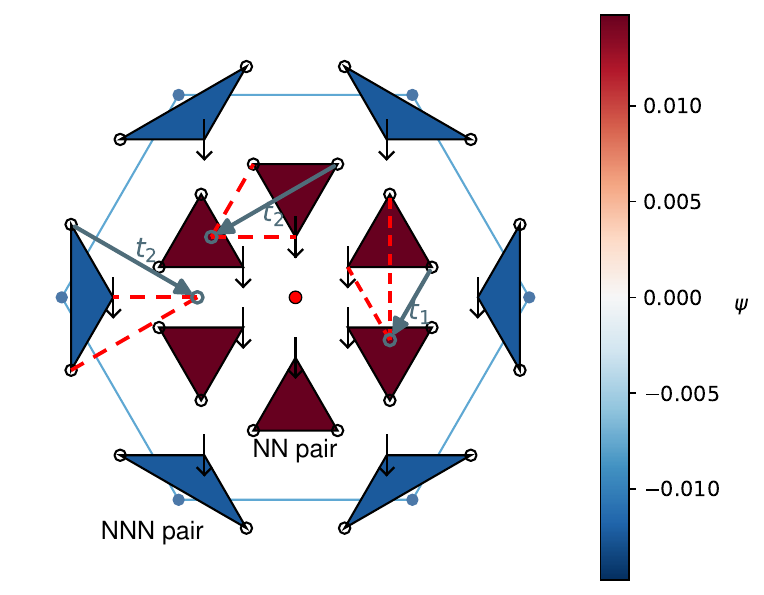} %
  \caption{The state map of the six NN and six NNN hole
pairs (located at the blue markers) surrounding a fixed magnon at the center (red marker)
reveals that the pairing manifold intrinsically possesses $s$–wave symmetry. Moreover,
the next–nearest–neighbor hopping $t_{2}$ enables transitions between symmetry–related NN
configurations as well as between symmetry–related NNN configurations, while the
nearest–neighbor hopping $t_{1}$ connects a NN pair to a NNN pair. 
}
  \label{fig:2}
\end{figure}

To illustrate this symmetry, we focus on the two geometries with the largest amplitudes in Fig.~\ref{fig:1}: 
(i) a nearest-neighbor (NN) configuration in which the two holes and the magnon are mutually at nearest-neighbor separation, and 
(ii) a next-nearest-neighbor (NNN) configuration in which the holes are next-nearest neighbors, while each remains at nearest-neighbor separation from the magnon. 
Fixing the magnon at $(0,0)$ and placing the holes on the corresponding sites, we compute the wave function under lattice rotations. 
As shown in Fig.~\ref{fig:2}, both configurations transform trivially under these rotations, confirming the $s$-wave character of the bound state.

The connectivity of the pairing geometries is governed by the hopping processes. For the dominant NN configuration, states related by a $C_{6}$ rotation of $\pi/3$ are connected by a single $t_{2}$ hop of one hole; likewise, NNN configurations related by a $C_{3}$ rotation of $2\pi/3$ are also linked by a single $t_{2}$ hop. Nearest-neighbor hopping $t_{1}$ further connects NN and NNN geometries. Consequently, the entire pairing manifold is connected through combinations of $t_{1}$ and $t_{2}$ processes, promoting coherent superposition of symmetry-related states, lowering the total energy, and stabilizing a finite binding energy,
\begin{equation}
E_b \equiv E(1h1m)+E(1h)-E(2h1m)\simeq0.30,
\label{eq:binding_energy}
\end{equation}
for $t_{1}=1$ and $t_{2}=0.3$.

In the Supplemental Material (SM) we compute the $2h1m$ dispersion using the Lippmann--Schwinger equation and extract the corresponding effective mass. For $t_{2}=0.3$, the effective mass $m^{*}$ is substantially reduced compared to $t_{2}=0.16$, enhancing the mobility of the bound pairs. In the dilute limit, essentially all composites participate in the condensate, so that the superfluid density satisfies $n_s \simeq \delta/2$ (density of hole pairs). The reduced effective mass therefore leads to an enhanced phase stiffness, $\rho_s \sim (\delta/2)/m^{*}$, favoring SC. 

Motivated by these observations, we perform DMRG simulations using ITensor~\cite{itensor} for finite hole density $\delta$ at $t_1=1,t_{2}=0.3$. We study a $24\times4$ cylindrical geometry with open boundaries along $\hat{x}$ and compute the magnetization $M$ as a function of $h$ for several values of $\delta$ at low temperature $k_BT=0.03t_1$. The calculations employ a maximum bond dimension of $8000$ and a truncation cutoff of $10^{-6}$.

\begin{figure}[t] %
  \centering
  \includegraphics[width=1\columnwidth]{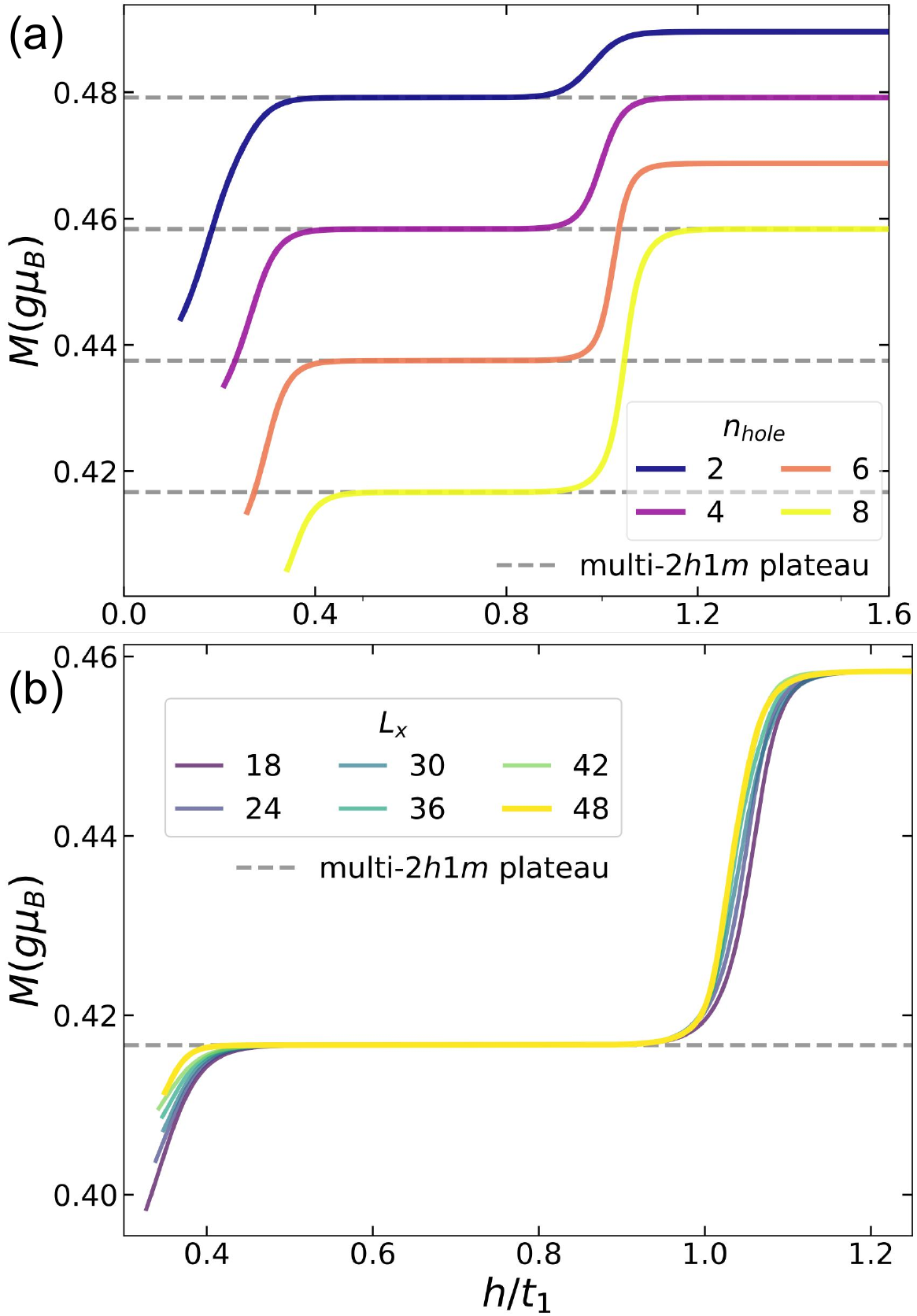} 
  \caption{Magnetization $M$ as a function of magnetic field $h$ on four-leg cylinders at low temperature $k_BT=0.03t_1$: (a) for several number of hole dopings up to $n_{hole}=8$ at fixed length $L_x=24$, and (b) for system lengths up to $L_x=48$ at fixed doping $\delta=1/12$. In both cases, a finite field window $0.5t_1 < h < t_1$ is observed in which the system is stabilized in the multi-$2h1m$ bound-state regime.}
  \label{fig:3}
\end{figure}

Upon lowering the magnetic field $h$ from above saturation, we observe a broad magnetization plateau at incomplete polarization (see Fig.~\ref{fig:3}). 
By comparing the measured magnetization with the value expected for a dilute gas of $2h1m$ quasiparticles, $M_p=(1-2\delta)/2$, we identify this plateau as a phase stabilized by a finite density of $2h1m$ bound states [Fig.~\ref{fig:3}(a)]. 
Fixing the hole density to $\delta=1/12$ and varying the system length $L_x$ on an $L_x\times4$ cylindrical geometry, we find that the resulting magnetization curves $M(h)$ are essentially independent of $L_x$ [Fig.~\ref{fig:3}(b)].  This robustness establishes a controlled starting point for exploring unconventional superconductivity emerging from this phase.

\paragraph{}

\paragraph{}

The primary diagnostic of superconducting order is the SC pair--pair correlation function~\cite{ChenHaldaneSheng_PNAS_2025_tJ_Global,JiangDevereauxJiang_PRB_2024_6leg,GongZhuSheng_PRL_2021_tJ_RobustDwave,JiangScalapinoWhite_PNAS_2021_ttprimeJ,JiangZaanenDevereauxJiang_PRResearch_2020_4leg,JiangChenWahlDevereauxKivelson_Science_2019,JiangKivelson_ProcNAS_2022_StripeEnhanced}, defined as
\begin{equation}
\Phi(r)_{\hat{\mathbf{r}}_1,\hat{\mathbf{r}}_2,\hat{\mathbf{r}}_3,\hat{\mathbf{r}}_4}
\equiv
\big\langle
\hat{\Delta}^{\dagger}_{\hat{\mathbf{r}}_1,\hat{\mathbf{r}}_2,(x,y)}
\hat{\Delta}^{\;}_{\hat{\mathbf{r}}_3,\hat{\mathbf{r}}_4,(x+r,y)}
\big\rangle .
\end{equation}
Each three-body pair operator
$\hat{\Delta}^{\dagger}_{\hat{\mathbf{r}}_1,\hat{\mathbf{r}}_2,(x,y)}$
and
$\hat{\Delta}^{\;}_{\hat{\mathbf{r}}_3,\hat{\mathbf{r}}_4,(x+r,y)}$
represents a triangular unit composed of two holes and one magnon, while $r$
denotes the separation between the two triangular units along the $\hat{x}$
direction. To suppress boundary effects, 
we choose the reference position at
$x=L_x/4$ and $y=L_y/2$.

In 1D systems, true long-range superconducting order is ruled out by the Mermin--Wagner theorem~\cite{MerminWagner1966}. Quite generally, only quasi-long-range order is possible, with the pair--pair correlations decaying as a power-law with distance $r$:
$\Phi(r)\sim  r^{-K_{\mathrm{SC}}}$. We focus on translationally related pair--pair correlations built from 
(i) a pair of parallel NN pairs with displacement vectors 
$\hat{\mathbf{r}}_1=\hat{\mathbf{r}}_3=(1,0)$ and 
$\hat{\mathbf{r}}_2=\hat{\mathbf{r}}_4=\bigl(\tfrac{1}{2},-\tfrac{\sqrt{3}}{2}\bigr)$, 
and (ii) a pair of parallel NNN pairs with 
$\hat{\mathbf{r}}_1=\hat{\mathbf{r}}_3=(1,0)$ and 
$\hat{\mathbf{r}}_2=\hat{\mathbf{r}}_4=\bigl(-\tfrac{1}{2},-\tfrac{\sqrt{3}}{2}\bigr)$.
Using DMRG, we solve the Hubbard model with the magnetic field tuned to the multi-$2h1m$ magnetization plateau
and study a $L_{x}\times 6$ cylinder with $L_{x}=32$ at doping $\delta=1/12$ (maximum bond dimension $m=15{,}000$) and a $L_{x}\times 4$ 
cylinder at $\delta=1/8$ ($m=12{,}000$). Convergence of  numerical results as a function of bond dimension is shown in the SM.

The resulting $\Phi(r)$ shown in 
Figs.~4(a,b) exhibit power-law decay $\Phi(r)\sim r^{-K_{\mathrm{sc}}}$, with exponents $K_{\mathrm{sc}}\simeq1.37$ and $1.33$ for NN and NNN pairs on the six-leg cylinder at $\delta=1/12$, well as $K_{\mathrm{sc}}\simeq1.70$ and $1.68$ for NN and NNN pairs on the four-leg cylinder at $\delta=1/8$. This behavior is characteristic
of a Luther--Emery (spin-gapped) regime in quasi-1D, wherein superconducting correlations display power-law decay while transverse spin correlations are short-ranged~\cite{LutherEmeryPRL1974,HaldaneJPhysC1981,VoitRPP1995,GiamarchiBook2004}. In all cases $K_{\mathrm{sc}}<2$, implying a superconducting susceptibility 
$\chi_{\mathrm{sc}}\sim T^{-(2-K_{\mathrm{sc}})}$~\cite{GiamarchiBook2004,VoitRPP1995} that diverges as $T\!\to\!0$, consistent with the quasi--long-range superconducting order. The near equality of $K_{\mathrm{sc}}$ for NN and NNN pairs within the same system further suggests a unified quasi--long--range order even when the Cooper pairs possess real-space dispersion.

We also examine pair--pair correlations constructed solely from a two-hole operator,
\begin{equation}
\Phi_{2h}(r)\equiv 
\big\langle \Delta^{\dagger}_{(x,y)}\,\Delta_{(x+r,y)}\big\rangle ,
\label{eq:phi2h_def}
\end{equation}
where $\Delta_{(x,y)}\equiv c_{(x,y)\uparrow}\,c_{(x,y)+\hat{\mathbf e}\uparrow}$ creates a nearest-neighbor hole pair with bond vector $\hat{\mathbf e}=(-\tfrac{1}{2},\tfrac{\sqrt{3}}{2})$. To minimize boundary effects, we fix the reference position at $x=L_x/4$, $y=L_y/2$ and evaluate $\Phi_{2h}(r)$ on a six-leg cylinder at $\delta=1/12$ and a four-leg cylinder at $\delta=1/8$, both with $L_x=32$.
As shown in Fig.~4(c), $\Phi_{2h}(r)$ decays exponentially, $\Phi_{2h}(r)\sim e^{-r/\xi_{\mathrm{sc}}}$, with correlation lengths $\xi_{\mathrm{sc}}\simeq1.21$ (six-leg, $\delta=1/12$) and $\xi_{\mathrm{sc}}\simeq1.43$ (four-leg, $\delta=1/8$). The short correlation length demonstrates the absence of long-range superconducting order arising from two-hole pairs alone, indicating that the magnon is an essential constituent of the Cooper pair rather than a mere mediator that induces attraction between holes. The condensate is formed not by holes, but by composite $2h1m$ quasiparticles.

Since the ground state corresponds to a dilute gas of holes and magnons doped into a spin polarized background, the system retains overall net magnetization. 
To characterize its magnetic correlations, we compute the transverse spin--spin correlation function
\begin{equation}
\begin{split}
F(r)\equiv {}&
\big\langle S^{x}(x,y)\,S^{x}(x{+}r,y)
+ S^{y}(x,y)\,S^{y}(x{+}r,y)\big\rangle ,
\end{split}
\label{eq:F_def}
\end{equation}
evaluated at $x=L_x/4$ and $y=L_y/2$, and restricted to the transverse ($x$ and $y$) spin components. For a six-leg cylinder at $\delta=1/12$ and a four-leg cylinder at $\delta=1/8$, the results shown in Fig.~4(d) exhibit exponential decay, $|F(r)|\sim e^{-r/\xi_s}$, with correlation lengths $\xi_s\simeq2.43$ (six-leg, $\delta=1/12$) and $\xi_s\simeq2.46$ (four-leg, $\delta=1/8$). The short-ranged nature of the transverse spin correlations reflects the competition between AFM tendencies and unconventional SC: by binding to hole pairs, magnons are prevented from condensing individually to form long-range AFM order.

\begin{figure}[t] %
  \centering
  \includegraphics[width=1\columnwidth]{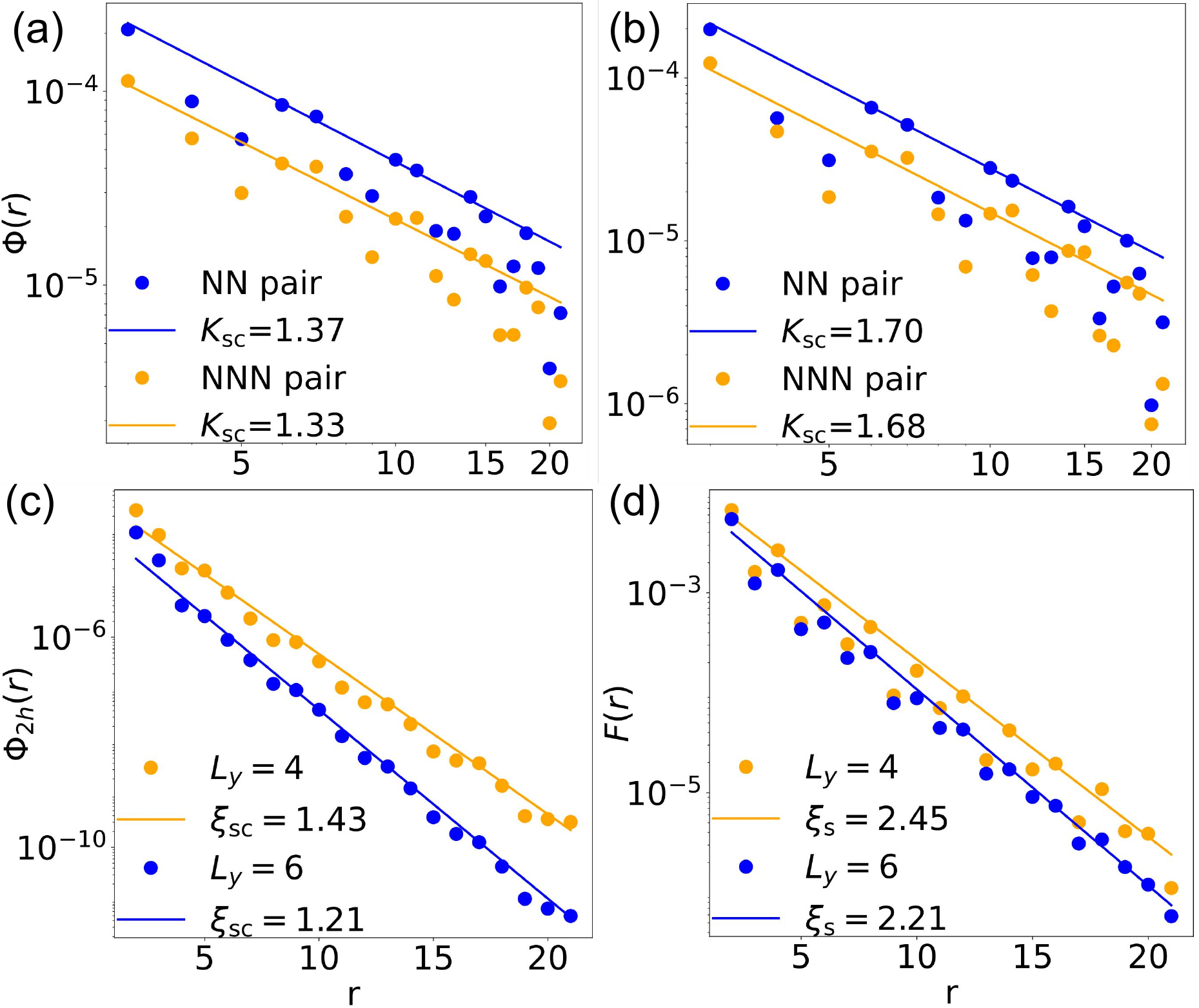} %
  \caption{ NN and NNN pair--pair correlations $\Phi(r)$ plotted on double-logarithmic scales in (a) and (b) for the $N=6\times32,\ \delta=1/12$ (bond dimension $m=15000$) and $N=4\times32,\ \delta=1/8$ (bond dimension $m=12000$) cylinders, respectively. In both cases, $\Phi(r)$ follows a power-law decay, $\Phi(r)\sim r^{-K_{\mathrm{sc}}}$. In contrast, the $2h$-only pair correlations $\Phi_{2h}(r)$, shown on a semi-logarithmic scale in (c), decay exponentially, $\Phi_{2h}(r)\sim e^{-r/\xi_{\mathrm{sc}}}$. The transverse spin--spin correlation $F(r)$, plotted on a semi-logarithmic scale in (d), also exhibits exponential decay, $F(r)\sim e^{-r/\xi_{\mathrm{s}}}$.}
  \label{fig:4}
\end{figure}
\paragraph{}

From an experimental perspective, realizing magnon-mediated superconductivity in moir\'e or cold-atom Hubbard platforms would be highly significant, providing the first direct evidence of unconventional superconductivity emerging from strong repulsive interactions. As in the experimental observation of the predicted gas of $S=3/2$ polarons~\cite{ZhangZhuBatista2018,ZhangFu2023,Tao2024SpinPolaron}, the magnetization plateau identified in this work offers a powerful diagnostic for characterizing this new state of matter. A key experimental challenge is to engineer a sizable next-nearest-neighbor hopping $t_2 \sim 0.3 t_1$, which optimizes the binding energy $E_b$ of the composite Cooper pairs. The value $E_b=0.3t_1$ obtained in Eq.~\eqref{eq:binding_energy} is comparable to the binding energy $\simeq 0.5 t_1$ of the $S=3/2$ polarons in the absence of $t_3$~\cite{ZhangZhuBatista2018}. This suggests that the magnetization plateau reported here should become observable at temperatures comparable to those at which the $S=3/2$ polaron plateau was detected in moir\'e heterostructures~\cite{Tao2024SpinPolaron,Ciorciaro_Nature_2023}. We further propose a targeted search in moir\'e homobilayers by sweeping the displacement field to modulate $t_2/t_1$ and stabilize the $2h1m$ state at intermediate doping.

As briefly illustrated in the Supplemental Material, the interplay between magnon-mediated superconductivity and competing charge- and spin-density-wave (CDW and SDW) tendencies warrants further investigation. More broadly, the unconventional pairing mechanism uncovered here can be generalized to a wide class of strongly correlated frustrated systems, offering a new route toward unconventional superconductivity beyond traditional paradigms.

\begin{acknowledgments}
\textbf{Acknowledgments} We thank Dr. Muqing Xu and Nianlong Zou for helpful discussions. 
Computations were performed on the group nodes and UTK ISAAC cluster. This work was supported by Max Planck partner lab grant on quantum materials.
C.D.B. acknowledges support from the U.S. Department of Energy, Office of Science, Office of Basic Energy Sciences, under Award Number DE-SC0022311. 
\end{acknowledgments}

\bibliographystyle{apsrev4-2}
\bibliography{refs}

% ...

\end{document}


\title{Supplemental Material for \\ 
 ``Magnon-Mediated Superconductivity in the Infinite-$U$ Triangular Lattice''}

\author{Hantian Zhu}
\affiliation{Department of Physics and Astronomy, University of Tennessee, Knoxville, Tennessee 37996, USA}

\author{Yixin Zhang}
\affiliation{Department of Physics and Astronomy, University of Tennessee, Knoxville, Tennessee 37996, USA}

\author{Shang-Shun Zhang}
\affiliation{Department of Physics and Astronomy, University of Tennessee, Knoxville, Tennessee 37996, USA}

\author{Yang Zhang}
\email{yangzhang@utk.edu}
\affiliation{Department of Physics and Astronomy, University of Tennessee, Knoxville, Tennessee 37996, USA}
\affiliation{Min H. Kao Department of Electrical Engineering and Computer Science, University of Tennessee, Knoxville, Tennessee 37996, USA}

\author{Cristian D. Batista}
\email{cbatist2@utk.edu}
\affiliation{Department of Physics and Astronomy, University of Tennessee, Knoxville, Tennessee 37996, USA}
\affiliation{Neutron Scattering Division, Oak Ridge National Laboratory, Oak Ridge, TN, USA}

\date{\today}
\maketitle

\setcounter{equation}{0}
\setcounter{figure}{0}
\setcounter{table}{0}
\renewcommand{\theequation}{S\arabic{equation}}
\renewcommand{\thefigure}{S\arabic{figure}}
\renewcommand{\thetable}{S\arabic{table}}

\section{Effective mass of the $2h1m$ pair}
\label{app:eff_mass}

\begin{figure}[t]
\centering
\includegraphics[width=0.7\textwidth]{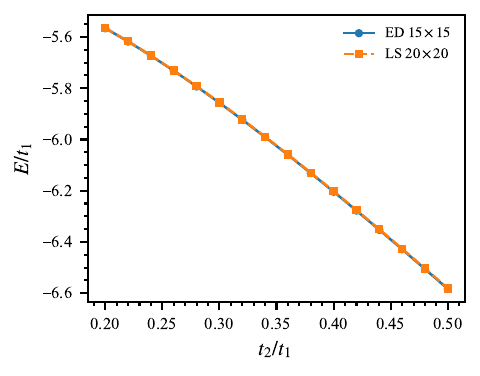}
\caption{Ground-state energy $E$ of the $2h1m$ bound state as a function of the
next-nearest-neighbor hopping $t_2$, obtained from exact diagonalization (ED) on a $15\times15$ lattice (circles) and from the Lippmann--Schwinger (LS) equation evaluated on a $20\times20$ momentum grid (squares). The two methods show excellent agreement over the entire range of $t_2$.}
\label{fig:s1}
\end{figure}

To quantify the mobility of the composite $2h1m$ pair, we compute its dispersion using the Lippmann--Schwinger (LS) equation and extract the corresponding effective mass. The wave function for two holes and one magnon can be expressed as follows:

\begin{equation}
\lvert\psi_{\mathbf Q}\rangle
= \sum_{\mathbf{r}_1,\mathbf{r}_2} \psi_{\mathbf{Q}}(\mathbf{r}_1,\mathbf{r}_2)\,
\frac{1}{\sqrt{N}} \sum_{\mathbf{R}} e^{i\mathbf{Q}\cdot\mathbf{R}}
\lvert \mathbf{R},\, \mathbf{R}+\mathbf{r}_1,\, \mathbf{R}+\mathbf{r}_2 \rangle .
\tag{S1}
\end{equation}
Here $\mathbf{R}$ is the position vector of the mobile magnon, $\mathbf{r}_{1,2}$ is the relative position vector of two holes with respect to the magnon, and $N$ is the  number of lattice sites. The fermionic statistics of holes implies $\psi_{\mathbf{Q}}(\mathbf{r}_1,\mathbf{r}_2)=-\psi_{\mathbf{Q}}(\mathbf{r}_2,\mathbf{r}_1)$. The $2h1m$ wave function with total momentum  $\mathbf{Q}$ can be Fourier transformed as a function of the relative position $\mathbf{r}$ of one hole with respect to the magnon and the relative momentum $\mathbf{p}$ of the other hole with respect to the magnon:
\begin{equation}
    \psi_{\mathbf{Q}}(\mathbf{p}, \mathbf{r}) = \frac{1}{\sqrt{N}} \sum_{\mathbf{\mathbf{r}}^{\prime}} e^{-i \mathbf{p} \cdot \mathbf{r}^{\prime}} \, \psi_{\mathbf{Q}}(\mathbf{r}^{\prime}, \mathbf{r}).
\end{equation}
From this wavefunction representation of the three-body state, we obtain the Lippmann–Schwinger equation. First, we separate the Hamiltonian into the free-particle part and the interaction part:
\begin{align}
    \mathcal{H} =\mathcal{H}_0+ \mathcal{H}_{\rm int}.
\end{align}
The free-particle Hamiltonian is:
\begin{align}
    \mathcal{H}_0 = \sum_{\mathbf{p}} \xi_\mathbf{p} c_\mathbf{p}^\dagger c_\mathbf{p} + \sum_\mathbf{q} \omega_\mathbf{q} b_\mathbf{q}^\dagger b_\mathbf{q},
\end{align}
where $c_{p}^{\dagger}$ creates a hole with momentum $p$ and $b_{q}^{\dagger}$ creates a magnon with momentum $q$. Since the magnon is  immobile in the infinite–$U$ Hubbard limit, $\omega_\mathbf{q}=0$.
We consider the triangular–lattice dispersion for nearest-neighbor hopping $t_{1}$ and next nearest neighbor hopping $t_{2}$. After setting the lattice constant $a=1$, and choosing the lattice vectors  $\mathbf{a}_{1}=(1,0)$, $\mathbf{a}_{2}=(\tfrac12,\tfrac{\sqrt{3}}{2})$, we obtain the following expression for the dispersion relation:
\begin{equation}
\xi_{\mathbf{p}}
= 2t_{1}\!\left[\cos p_{x}
+2\cos\!\left(\tfrac{p_{x}}{2}\right)\cos\!\left(\tfrac{\sqrt{3}\,p_{y}}{2}\right)\right]
+2t_{2}\!\left[\cos(\sqrt{3}\,p_{y})
+2\cos\!\left(\tfrac{3p_{x}}{2}\right)\cos\!\left(\tfrac{\sqrt{3}\,p_{y}}{2}\right)\right].
\end{equation}
\begin{figure}[t]
\centering
\includegraphics[width=0.7\textwidth]{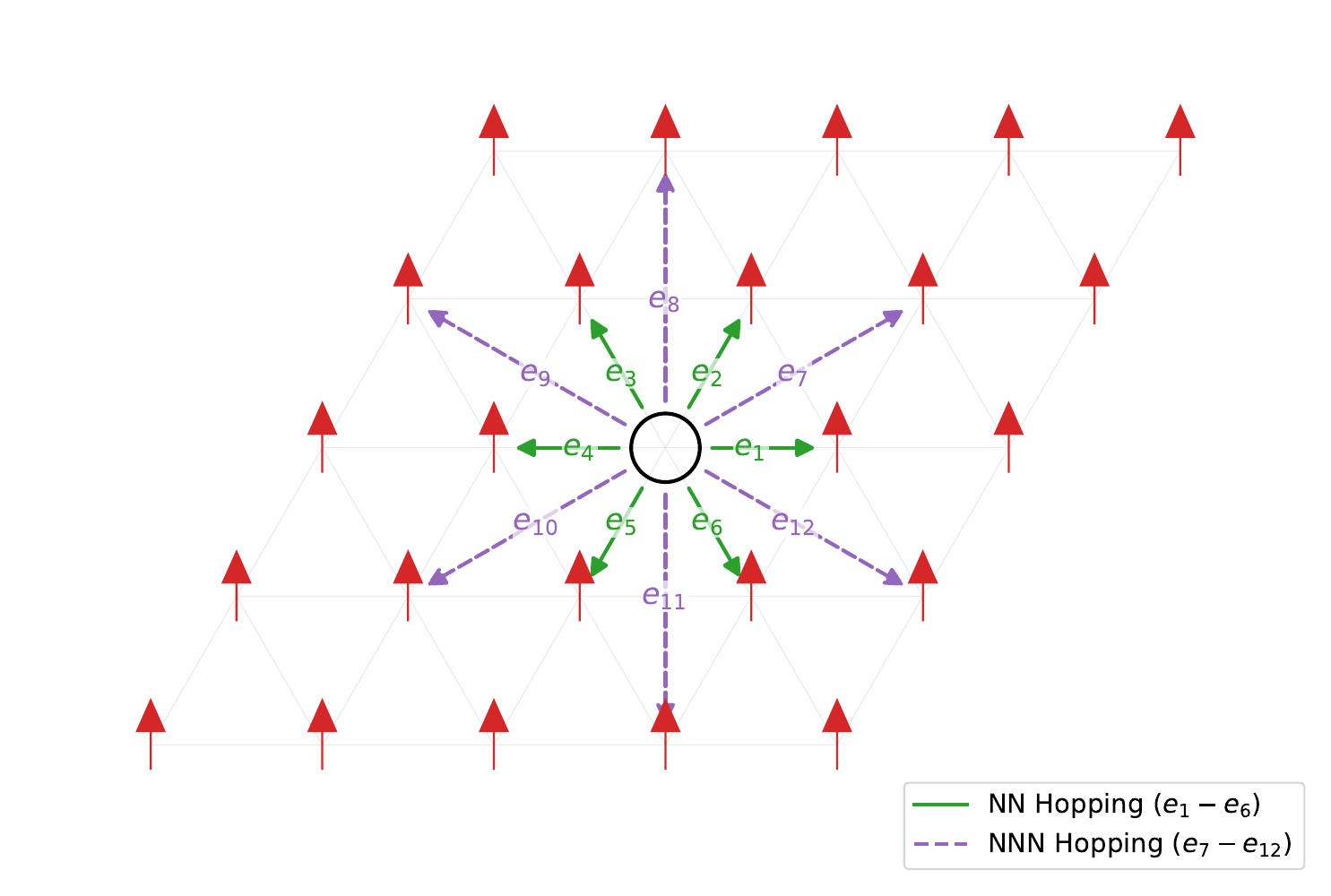}
\caption{The 6 nearest-neighbor hopping and 6 next-nearest hopping denoted by $\mathbf{e}_v$.}
\label{fig:hole_hopping}
\end{figure}
The interaction part of the Hamiltonian includes (i) processes in which the magnon hops from other sites to a site occupied by a hole, and (ii) an onsite energy term when the magnon and a hole reside on the same site:
\begin{align}
    H_{\rm int} = \sum_v t_v \hat{O}_v + V \sum_{\mathbf{r}} n_{\mathbf{r}} m_{\mathbf{r}}, 
\end{align}
where $\nu$ enumerates all the nearest and next-nearest neighbor bonds (see Fig.\ref{fig:hole_hopping}), $n_{\mathbf{r}} = 1-c_{\mathbf{r}}^\dagger c_{\mathbf{r}}$ and $m_{\mathbf{r}} = b_{\mathbf{r}}^\dagger b_{\mathbf{r}}$ are the number operators of the holes and magnons at site $\mathbf{r}$, and $V$ is taken as $\infty$ to enforce the hard-core constraint that forbids the magnon and a hole from occupying the same lattice site ($\mathbf r=0$ in the relative coordinate). The 
operator $\hat{O}_v$ 
acting on the quantum state basis $\lvert \mathbf{Q}; \mathbf{p}, \mathbf{r} \rangle$ is nonzero if and only if $\mathbf{r}$ equals to a nearest-neighbor or a next-nearest-neighbor hopping vector $\mathbf{e}_v$:

\begin{align}
    \hat{O}_v |\mathbf{Q}; \mathbf{p}, \mathbf{e}_v\rangle = e^{-i(\mathbf{Q}-\mathbf{p})\cdot \mathbf{e}_v} |\mathbf{Q}; \mathbf{p}, -\mathbf{e}_v\rangle.
\end{align}
In this hard-core limit ($V\to \infty$), the physical wave function satisfies the boundary condition $ \psi_{\mathbf{Q}}( \mathbf{p},\mathbf{r=0})=0$. Nevertheless, the product $V\,\psi_{\mathbf{Q}}( \mathbf{p},\mathbf{0})$ entering the LS equation remains finite: $\psi_{\mathbf{Q}}( \mathbf{p},\mathbf{0})$ scales as $1/V$ such that $V\psi_{\mathbf{Q}}( \mathbf{p},\mathbf{0})$ approaches a finite contact scattering amplitude, which is determined self-consistently by the LS equation.
For the bound state problem considered here, the 
LS equation reads as follows:
\begin{align}
    |\psi_{\mathbf Q}\rangle =  G_0^{(3)}(E) \mathcal{H}_{\text{int}} |\psi_{\mathbf Q}\rangle
\end{align}
where
\begin{align}
    G_0^
{(3)}(E) = \frac{1}{E - \mathcal{H}_0 + i\epsilon}.
\end{align}
We multiply $\langle \mathbf{Q}; \mathbf{p}, \mathbf{r} |$ from the left of both side, and get:
\begin{align}
    \psi_{\mathbf{Q}}(\mathbf{p}, \mathbf{r}) = \langle \mathbf{Q}; \mathbf{p}, \mathbf{r} | G_0^{(3)}(E) H_{\text{int}} |\psi_{\mathbf Q}\rangle.
\end{align}
Using the matrix elements of $H_{\rm int}$ under the quantum basis $\lvert \mathbf{Q}; \mathbf{p}, \mathbf{r} \rangle$ derived above, we get the integral formula of $\psi_{\mathbf{Q}}(\mathbf{p},\mathbf{r}) $:
\begin{align}
    \psi_{\mathbf{Q}}( \mathbf{p},\mathbf{r}) 
&= 2 \sum_v t_v \int \frac{d^2 \mathbf{k}}{(2\pi)^2} \, G^{(3)}_{\mathbf{Q}_E}(\mathbf{p},\mathbf{r}; \mathbf{k},-\mathbf{e}_v) \, e^{-i(\mathbf{Q}-\mathbf{k}) \cdot \mathbf{e}_v} \psi_{\mathbf{Q}}(\mathbf{k},\mathbf{e}_v) \nonumber \\
&\quad + 2V \int \frac{d^2 \mathbf{k}}{(2\pi)^2} \, G^{(3)}_{\mathbf{Q}_E}(\mathbf{p},\mathbf{r}; \mathbf{k}, \mathbf{0}) \psi_{\mathbf{Q}}(\mathbf{k},\mathbf{0}),
\end{align}
$G^{3}_{\mathbf{Q}_E}$ is the non-interacting three-body Green’s function
of the non-interacting system:
\begin{align}
G_{\mathbf{Q}_E}^{(3)}(\mathbf{p}_1, \mathbf{r}_1; \mathbf{p}_2, \mathbf{r}_2) 
&= \left\langle \mathbf{Q}; \mathbf{p}_1, \mathbf{r}_1 \left| \frac{1}{E - \mathcal{H}_0 + i\epsilon} \right| \mathbf{Q}; \mathbf{p}_2, \mathbf{r}_2 \right\rangle \nonumber \\
&= 2\pi^2 \delta_{\mathbf{p}_1 \mathbf{p}_2} \int \frac{d^2 \mathbf{k}}{(2\pi)^2} \frac{e^{-i{\mathbf k} \cdot (\mathbf{r}_2 - \mathbf{r}_1)}}{E - \left(\xi_{\mathbf{p}_2} + \xi_{\mathbf{k}} + \omega_{\mathbf{Q}-\mathbf{p}_2 - \mathbf{k}}\right)} \nonumber \\
&\quad - \frac{1}{2} \frac{e^{-i(\mathbf{p}_1 \cdot \mathbf{r}_2 - \mathbf{p}_2 \cdot \mathbf{r}_1)}}{E - \left(\xi_{\mathbf{p}_2} + \xi_{\mathbf{p}_1} + \omega_{\mathbf{Q}-\mathbf{p}_2 - \mathbf{p}_1}\right)}.
\end{align}
The hard-core constraint implies that at $\mathbf{r=0}$:
\begin{align}
0 
&= 2 \sum_v t_v \int \frac{d^2 \mathbf{k}}{(2\pi)^2} \, G_{QE}^{(3)}(\mathbf{p}, \mathbf{0}; \mathbf{k}, -\mathbf{e}_v) \, e^{-i(\mathbf{Q}-\mathbf{k}) \cdot \mathbf{e}_v} \psi_{\mathbf{Q}}(\mathbf{k}, \mathbf{e}_v) \nonumber \\
&\quad + 2V \int \frac{d^2 \mathbf{k}}{(2\pi)^2} \, G_{QE}^{(3)}(\mathbf{p}, \mathbf{0}; \mathbf{k}, \mathbf{0}) \psi_{\mathbf{Q}}(\mathbf{k}, \mathbf{0}).
\end{align}
The momentum integral over the first Brillouin zone in the LS equation is computed through the Gaussian quadrature method. Specifically, we replace the integral with 
by a discrete sum over Gaussian quadrature points,
\[
\int_{\text{BZ}} \frac{d^2\mathbf{k}}{(2\pi)^2}\, F(\mathbf{k})
\;\approx\;
\sum_{i,j = 1}^n w_{i} w_j\, F(k_x^i,k_y^j),
\]
where $w_i$ denotes the weight at each Gaussian quadrature point and we take $n=20$ in the calculations presented here. 
We  compare the resulting ground-state energy with exact diagonalization (ED) on a $15\times15$ lattice for several values of $t_2$. The LS energies are found to agree very well with the ED results over the entire parameter range considered as shown in the Fig.~\ref{fig:s1}, indicating that the LS treatment reliably captures the low-energy $2h1m$ bound state.

\begin{figure}[t]
\centering
\includegraphics[width=0.7\textwidth]{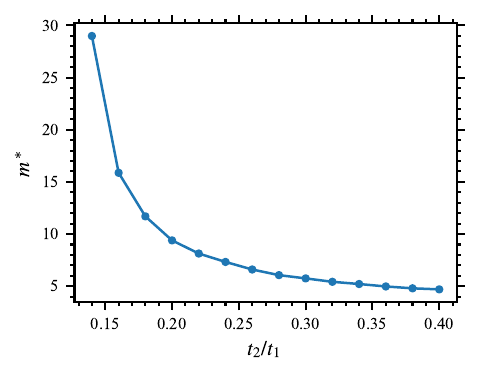}
\caption{Effective mass $m^\ast$ of the $2h1m$ pair as a function of the
next-nearest-neighbor hopping $t_2$ for $0.14 \le t_2/{\color{red} t_1} \le 0.40$.
The effective mass decreases rapidly when $t_2/{\color{red} t_1}$ is increased from $0.14$,
indicating that the composite pair becomes significantly lighter and more mobile
at larger $t_2$, which is favorable for the development of superconducting order. }
\label{fig:s2}
\end{figure}

We then vary the total momentum of the pair and obtain the full dispersion $E_{2h1m}(\mathbf{Q})$, from which the effective mass $m^\ast(t_2)$ is extracted from the curvature around the band minimum. Starting from $t_2 = 0.14t_1$, where the $2h1m$ bound state first becomes stable, and increasing $t_2$ up to $0.4t_1$, we find that $m^\ast$ initially decreases rapidly and then continues to decrease more slowly at larger $t_2$ as shown in Fig.~\ref{fig:s2}. The effective mass $m^\ast$ is expressed in units of $\hbar^2/(t_1 a^2)$, where $t_1$ is the nearest-neighbor hopping amplitude and $a$ is the lattice constant. This behavior shows that the composite pair becomes significantly lighter at sufficiently large $t_2$, and therefore can move more easily through the ferromagnetic background, providing a more favorable environment for superconducting coherence. In particular, the relatively small effective mass when $t_2 = 0.3t_1$ offers a natural rationale for focusing on this $t_2$ value when demonstrating the presence of quasi-long-range superconducting order in the main text.

\section{Charge and Spin Density Waves}

In quasi-one-dimensional geometries, there is a pronounced tendency toward
charge-density-wave (CDW) order. Indeed, DMRG studies of doped square-lattice
Hubbard models with next-nearest-neighbor hopping have shown that whenever
quasi-long-range superconducting correlations emerge, sizable CDW modulations
coexist~\cite{Jiang2025Competition}, particularly for small cylinder
circumferences \(L_y\). This behavior reflects, on the one hand, the presence of
Friedel oscillations induced by the open boundaries inherent to cylindrical
geometries, and, on the other hand, the enhanced one-dimensional character,
which suppresses superconducting correlations while favoring CDW order.

In our system, the situation is further enriched by the presence of both hole and magnon doping. The finite hole concentration naturally gives rise to charge oscillations and associated CDW patterns, while the finite density of magnons simultaneously induces oscillations in the longitudinal spin component $S^z$, leading to pronounced spin-density-wave (SDW) behavior. In this section, we analyze these numerical signatures of CDW and SDW order in detail.

\begin{figure}[t]
\centering
\includegraphics[width=0.7\textwidth]{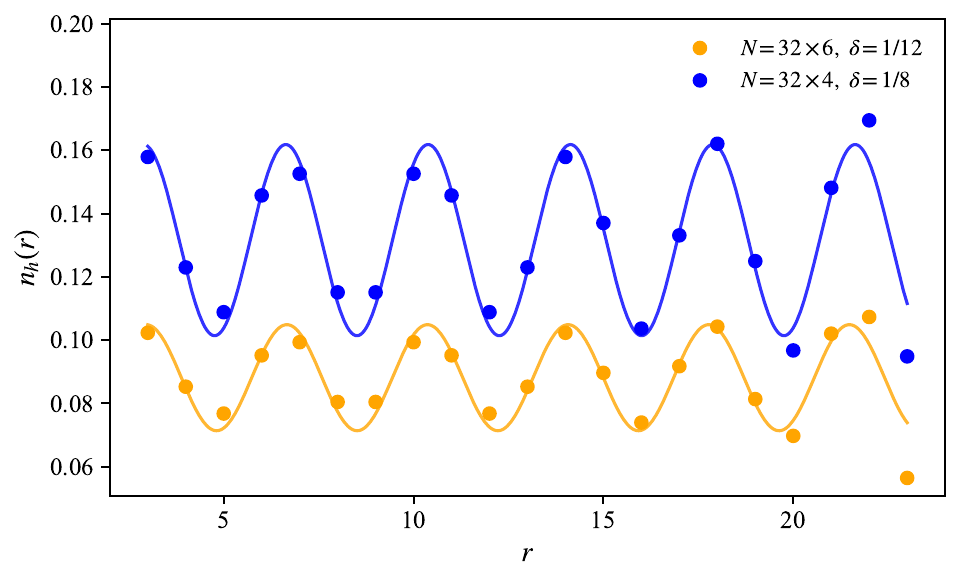}
\caption{Hole density profile $n_h(r)$ along the $x$ direction on $L_x = 32$ cylinders with $L_y = 4$ (blue, $\delta=1/8$)
and $L_y = 6$ (orange, $\delta=1/12$).}
\label{fig:s3}
\end{figure}

\begin{figure}[t]
\centering
\includegraphics[width=0.7\textwidth]{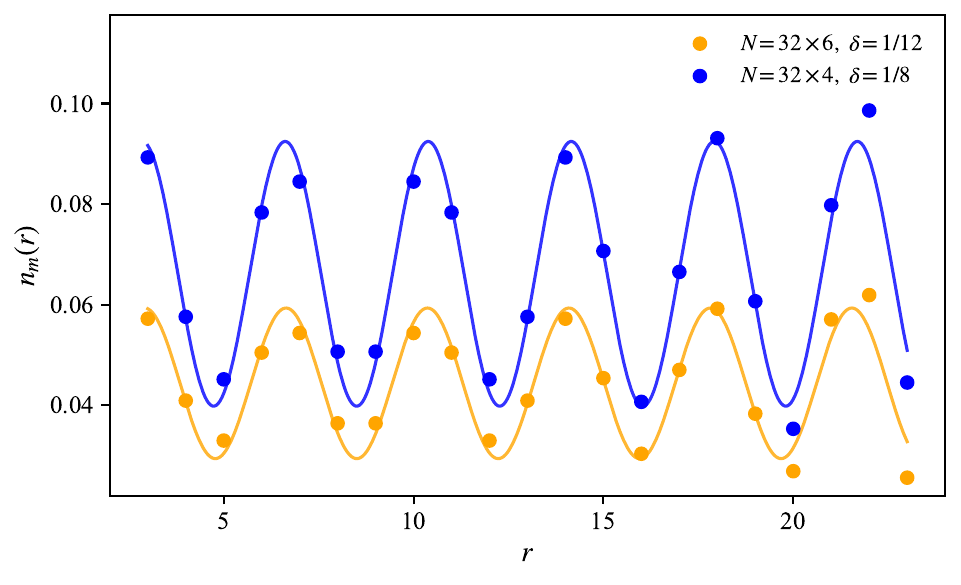}
\caption{Magnon density profile $ n_m(r)$ along the $x$ direction on
cylindrical geometries with $L_x = 32$ and $L_y = 4$ (blue, $\delta=1/8$)
and $L_y = 6$ (orange, $\delta=1/12$).}
\label{fig:s4}
\end{figure}

We first examine the spatial profiles of the hole and magnon densities. 
To this end, we consider one-dimensional cuts along the $x$ direction and define
\[
 n_h(r) = \big\langle 1- \hat n_{(x_0 + r, y_0)\uparrow} - \hat n_{(x_0 + r, y_0)\downarrow}\big\rangle, 
\qquad 
 n_m(r) = \big\langle \hat n_{(x_0 + r, y_0)\downarrow}\big\rangle,
\]
where we choose a reference point at $x_0 = L_x/4$ and $y_0 = L_y/2$ and calculate for a six--leg cylinder at $\delta=1/12$ and a four--leg cylinder at $\delta=1/8$
(both with $L_x=32$). The resulting hole density $n_h(r)$ and magnon density $ n_m(r)$ are shown in Fig.~\ref{fig:s3} and Fig.~\ref{fig:s4}, respectively.
Both quantities exhibit oscillations as a function of $r$. In particular, the oscillation amplitude grows significantly as $x_0+r$ approaches the open boundary, reflecting the strong boundary-induced modulations in this quasi-one-dimensional geometry.
Moreover, the dominant oscillation period is approximately four lattice spacings, consistent with a characteristic wavelength set by $L_y$ and the hole concentration $\delta$. For small $L_y$, the bound pairs tend to distribute rather uniformly along the $x$ direction, so that the density modulations naturally acquire a period of $2L_y/\delta$ in the $x$ direction for the parameters considered here. This provides a simple explanation for the strong density oscillations observed along the cylinder, even in the presence of quasi-long-range superconducting correlations.

\begin{figure}[t]
\centering
\includegraphics[width=0.7\textwidth]{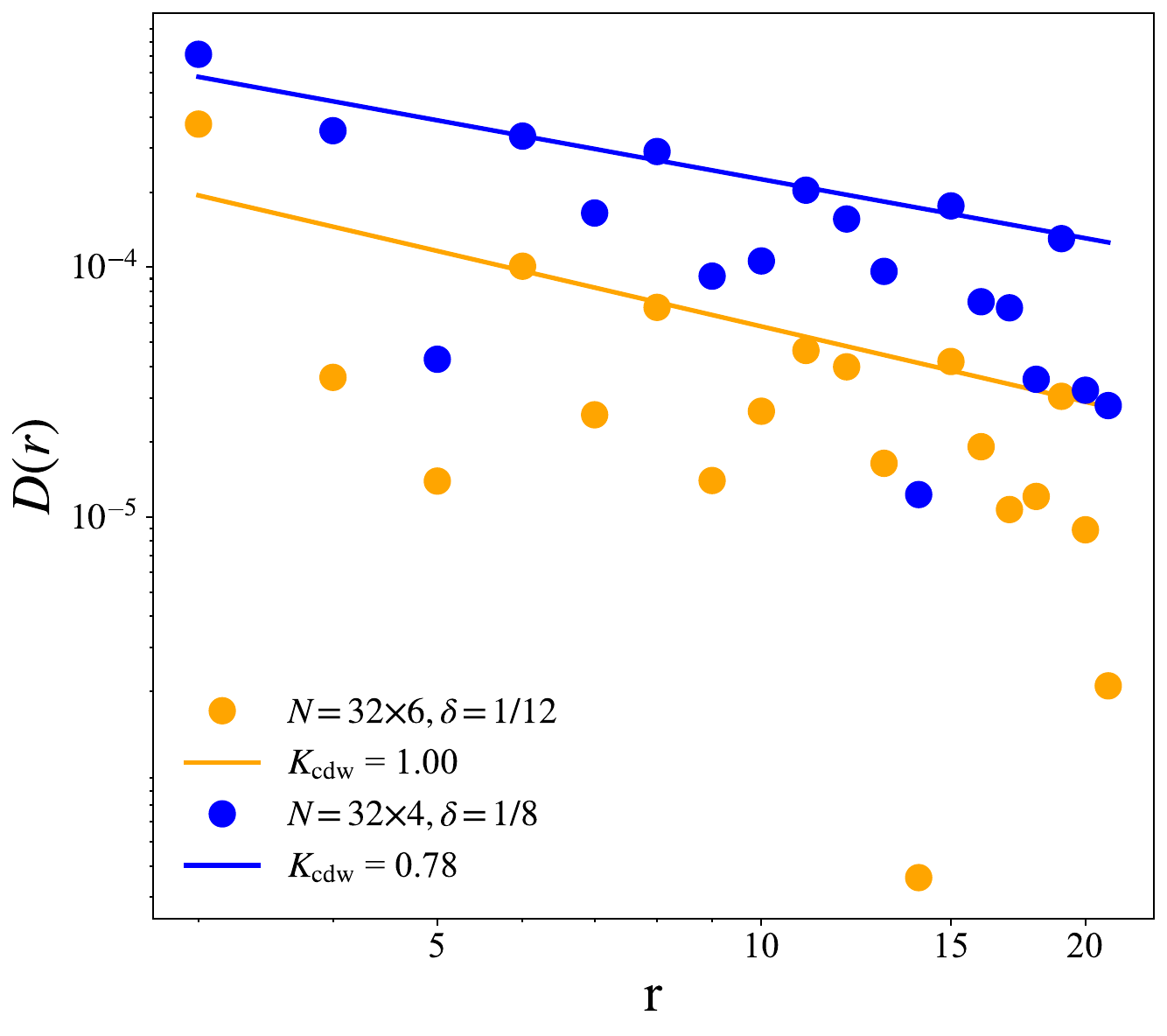}
\caption{Charge density-density correlation $D(r)$ plotted on double–logarithmic scales in (a) and (b) for the $N=6\times32,\delta=1/12$ (bond dimension $m=15000$)
and $N=4\times32,\delta=1/8$ (bond dimension $m=12000$) cylinders, respectively, both follow power–law decay
$\Phi(r)\sim r^{-K_{\mathrm{cdw}}}$}
\label{fig:s5}
\end{figure}

We also compute the charge density–density fluctuation correlation function
on both lattices, defined as
\begin{equation}
D(r)
= \big\langle
\big(\hat n_{(x_0,y_0)} - n_{(x_0,y_0)}\big)
\big(\hat n_{(x_0+r,y_0)} - n_{(x_0+r,y_0)}\big)
\big\rangle,
\label{eq:Dr_def}
\end{equation}
where $n_{x_0,y_0} = \langle \hat n_{(x_0,y_0)} \rangle$ and
$n_{(x_0+r,y_0)} = \langle \hat n_{(x_0+r,y_0)} \rangle$ denote the local average charge densities. As before, we fix the reference point at $x_0 = L_x/4$ and $y_0 = L_y/2$ and vary the separation $r$ along the $x$ direction. The resulting $D(r)$ exhibits a power-law decay with distance $r$:
$D(r)\sim  r^{-K_{\mathrm{cdw}}}$, as shown in Fig.~\ref{fig:s5}.
Comparing the CDW exponent \(K_{\mathrm{cdw}}\) extracted from \(D(r)\) with the
superconducting exponent \(K_{\mathrm{sc}}\) obtained in the main text, we find
that \(K_{\mathrm{sc}} > K_{\mathrm{cdw}}\) on both cylinders. However, upon
increasing the circumference from \(L_y=4\) to \(L_y=6\),
\(K_{\mathrm{sc}}\) decreases rapidly, whereas \(K_{\mathrm{cdw}}\) exhibits a
slight increase. This opposite trend suggests that, for wider cylinders with
\(L_y \sim 8\) or \(12\), the quasi-one-dimensional constraint may be further
relaxed, potentially leading to a regime in which superconducting correlations
dominate over CDW correlations.

On the other hand, the present model is purely kinetic-energy driven, with an
infinite on-site repulsion (Hubbard \(U=\infty\)). If additional spin
interactions of the \(t\)–\(J\) type are included, holes and magnons may further
lower the spin interaction energy by forming bound pairs. Indeed, existing
studies of the \(t\)–\(J\) model report robust quasi-long-range superconducting
order without accompanying CDW order~\cite{ChenHaldaneSheng_PNAS_2025_tJ_Global}.
In the main text, we have deliberately focused on magnon-mediated
superconductivity and its pairing mechanism as the central result of this work,
while a more systematic analysis of the competition and coexistence among
superconducting, CDW, and SDW orders is left for future investigation.

\begin{figure}[t]
\centering
\includegraphics[width=0.7\textwidth]{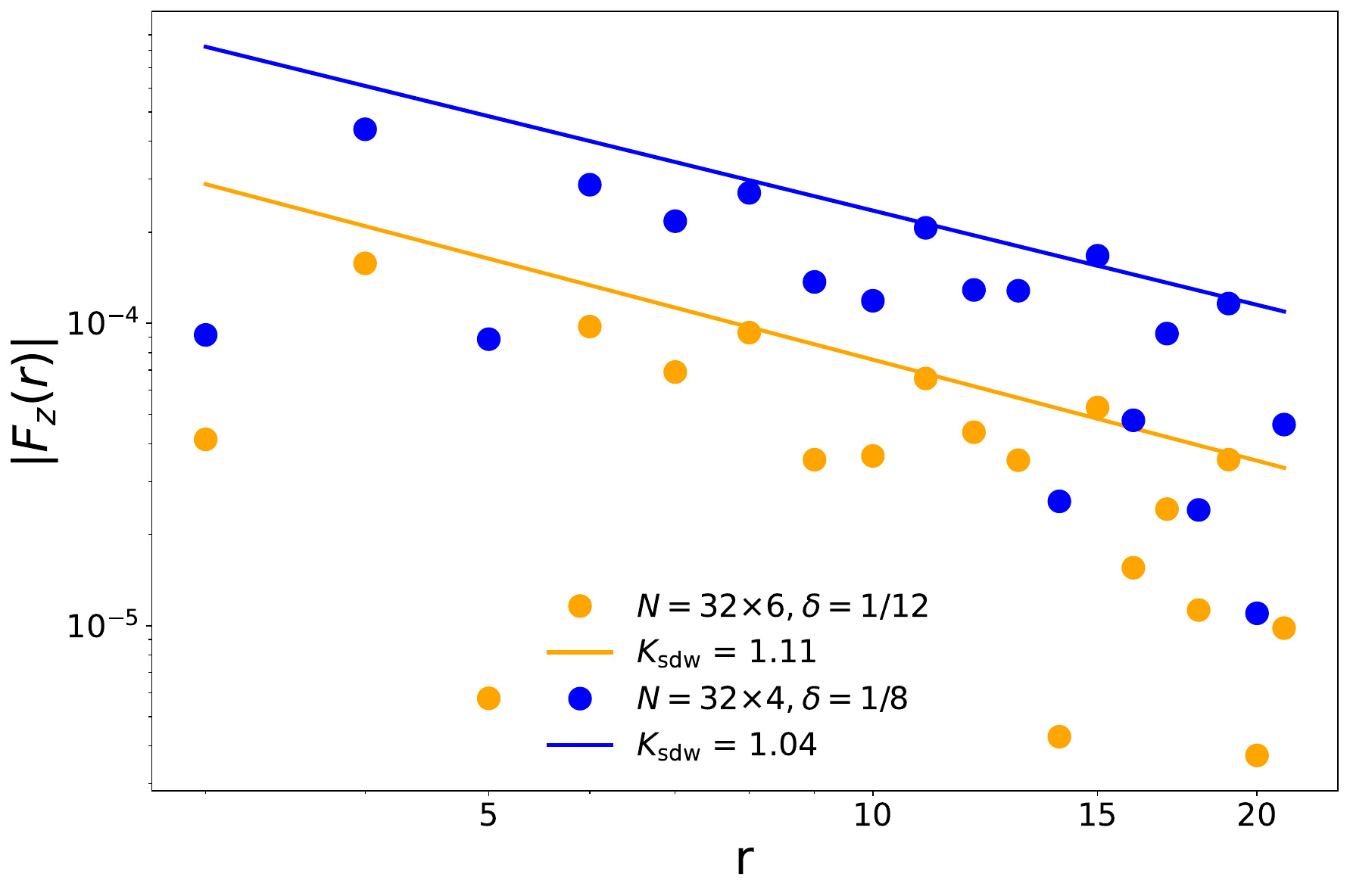}
\caption{The \(z\)-component of the spin--spin correlation function \(|F_z(r)|\)
plotted on double--logarithmic scales in (a) and (b) for the
\(N=6\times 32,\, \delta=1/12\) (bond dimension \(m=15000\)) and
\(N=4\times 32,\, \delta=1/8\) (bond dimension \(m=12000\)) cylinders,
respectively. In both cases, the data follow a power-law decay
\(|F_z(r)| \sim r^{-K_{\mathrm{sdw}}}\).}
\label{fig:s6}
\end{figure}

We similarly compute the longitudinal spin-density-wave (SDW) correlation
function along the \(z\) direction, defined as
\begin{equation}
F_z(r)
= \big\langle
\big(\hat S^z_{(x_0,y_0)} - S^z_{(x_0,y_0)}\big)
\big(\hat S^z_{(x_0+r,y_0)} - S^z_{(x_0+r,y_0)}\big)
\big\rangle,
\label{eq:Sz_def}
\end{equation}
where \(S^z_{(x_0,y_0)} = \langle \hat S^z_{(x_0,y_0)} \rangle\) and
\(S^z_{(x_0+r,y_0)} = \langle \hat S^z_{(x_0+r,y_0)} \rangle\) denote the local
average magnetization along the \(z\) direction. As in the charge sector, we fix
the reference point at \(x_0 = L_x/4\) and \(y_0 = L_y/2\), and vary the separation
\(r\) along the \(x\) direction. The resulting \(F_z(r)\) exhibits a power-law
decay with distance \(r\), \(|F_z(r)| \sim r^{-K_{\mathrm{sdw}}}\), as shown in
Fig.~\ref{fig:s6}.

We find that the longitudinal SDW correlations \(F_z(r)\) exhibit behavior very
similar to that of the CDW correlations \(D(r)\) as a function of distance
\(r\). This similarity indicates that, in the present system, magnons play a
role analogous to that of holes.

\section{Details of the DMRG calculation}

In this section, we briefly discuss technical aspects of the DMRG ground-state
calculations. We employ the DMRG algorithm as implemented in the
\texttt{ITensor} library~\cite{itensor} on finite cylindrical geometries.
Compared with conventional doped Hubbard or \(t\)–\(J\) models with nearest-
and next-nearest-neighbor hopping on the square lattice, our multi-\(2h1m\)
system involves a very low density of magnons relative to the fully polarized
background: the number of spin-down electrons is of order \(\delta/2\), such
that \(N_\uparrow \gg N_\downarrow\). By contrast, most DMRG studies of
superconductivity in doped Hubbard or \(t\)–\(J\) models on square lattices are
performed in the total-\(S^z=0\) sector, starting from an antiferromagnetic
parent state with \(N_\uparrow \simeq N_\downarrow\). As a consequence, the
Hilbert subspace explored in our simulations is substantially smaller than in
the usual AFM-based setups.

Furthermore, we work in the infinite-\(U\) limit and explicitly forbid double
occupancy, which further reduces the accessible Hilbert space. We study an
\(L_x \times 6\) cylinder with \(L_x = 32\) at doping \(\delta = 1/12\) using a
maximum bond dimension of \(m = 15{,}000\), and an \(L_x \times 4\) cylinder at
\(\delta = 1/8\) with \(m = 12{,}000\). For these parameters, the chosen bond
dimensions are sufficient: the final DMRG truncation errors are as small as
\(3.61 \times 10^{-8}\) for the six-leg cylinder at \(\delta = 1/12\) and
\(1.92 \times 10^{-9}\) for the four-leg cylinder at \(\delta = 1/8\).

\begin{figure}[t]
\centering
\includegraphics[width=0.7\textwidth]{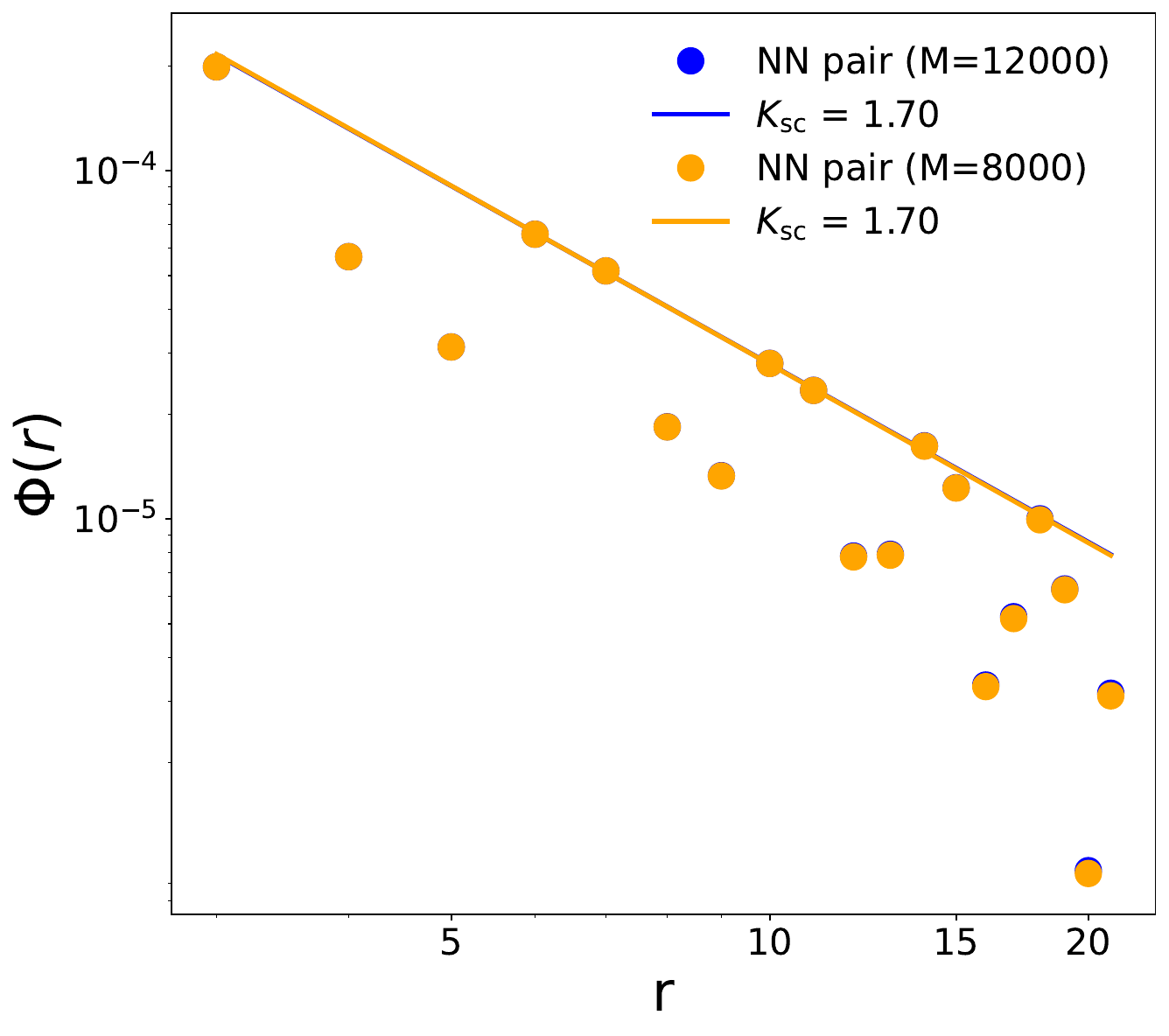}
\caption{The nearest-neighbor pair–pair correlation function
on an $L_x=32$ four-leg cylinder at doping $\delta = 1/8$ for two different
maximum bond dimensions, $m=12{,}000$ and $m=8{,}000$.
The two datasets are almost identical over the entire distance range,
demonstrating that the DMRG results are well converged with respect to
the bond dimension.}
\label{fig:s8}
\end{figure}

\begin{figure}[t]
\centering
\includegraphics[width=0.7\textwidth]{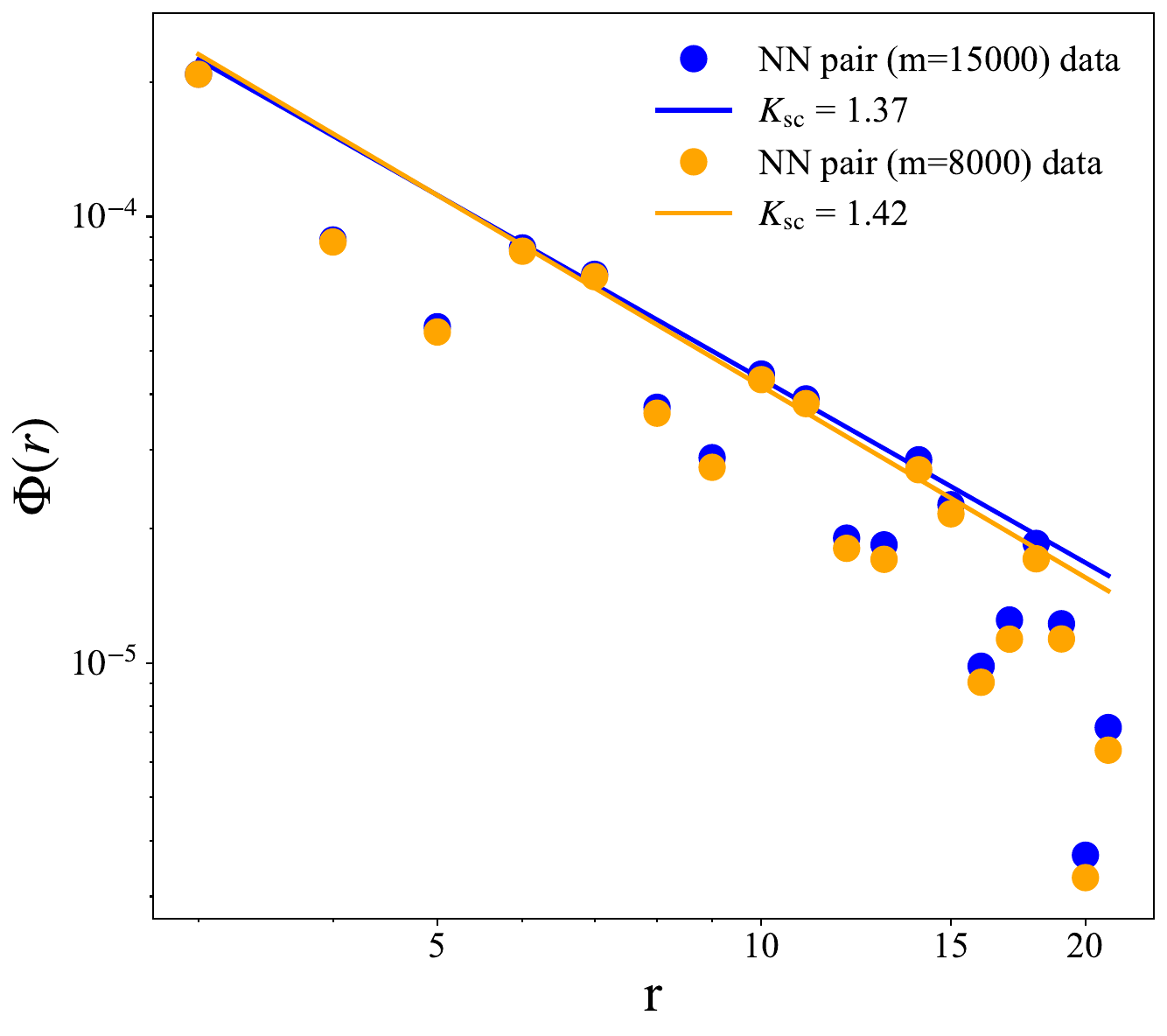}
\caption{The nearest-neighbor pair–pair correlation function
on an $L_x=32$ six-leg cylinder at doping $\delta = 1/12$ for two different
maximum bond dimensions, $m=15{,}000$ and $m=8{,}000$.
The two datasets are almost identical over the entire distance range, demonstrating that the DMRG results are well converged with respect to the bond dimension.}
\label{fig:s7}
\end{figure}

We have also verified convergence with respect to the bond dimension \(m\) for
both lattice geometries. In addition to the largest bond dimensions quoted
above, we performed DMRG calculations with a reduced bond dimension of
\(m = 8000\) and evaluated the nearest-neighbor superconducting pair--pair
correlation functions using the same procedure as in the main text. As shown in
Figs.~\ref{fig:s8} and \ref{fig:s7}, the resulting pair--pair correlators for
\(L_x \times 4\) at \(\delta = 1/8\) and \(L_x \times 6\) at \(\delta = 1/12\) are
almost indistinguishable from those obtained with \(m = 12{,}000\) and
\(m = 15{,}000\), respectively. This demonstrates that our DMRG simulations are
well converged with respect to the bond dimension and that the extracted
quasi-long-range superconducting correlations are numerically robust.

\bibliographystyle{apsrev4-2}
\bibliography{refs}